\documentclass[]{aa}
\usepackage{aalongtable}
\usepackage{graphicx}
\usepackage{txfonts}
\usepackage{natbib}
\usepackage{balance}
\usepackage[figuresright]{rotating}
\usepackage[a4paper,breaklinks]{hyperref}
\idline{10}{10}

\begin{document}
\def\teff{$\mathrm T_{eff }$ }
\def\logg {log\,g}
\def\lambo{$\lambda$ Boo }
\def\vsini {$v\,\sin i$ }
\def\kms {$\mathrm{km\, s^{-1}}$ }
\newcommand{\pun}[1]{\,#1}
\newcommand{\loggf}{\ensuremath{\log gf}}
\newcommand{\mlp}{\ensuremath{\alpha_{\mathrm{MLT}}}}
\newcommand{\LHD}{{\sf LHD}}
\newcommand{\xx}{\ensuremath{\mathrm{1D}_{\mathrm{LHD}}}}
\newcommand{\cobold}{{\sf CO$^5$BOLD}}
\newcommand{\mD}{\ensuremath{\left\langle\mathrm{3D}\right\rangle}}
\newcommand{\linfor}{{\sf Linfor3D}}

\title{The solar photospheric abundance of carbon
}
    \subtitle{Analysis of atomic carbon lines with the CO5BOLD solar model }
\author{
E.~Caffau     \inst{1}
\and
H.-G.~Ludwig  \inst{2,1}
\and
P.~Bonifacio  \inst{2,1,3}
\and
R.~Faraggiana \inst{4}
\and
M.~Steffen    \inst{5}
\and
B.~Freytag    \inst{6}
\and
I.~Kamp       \inst{7}
\and
T.R.~Ayres    \inst{8}
 }

\institute{ 
GEPI, Observatoire de Paris, CNRS, Universit\'e Paris Diderot; 92195
Meudon Cedex, France
\and
CIFIST Marie Curie Excellence Team
\and
Istituto Nazionale di Astrofisica,
Osservatorio Astronomico di Trieste,  Via Tiepolo 11,
I-34143 Trieste, Italy
\and
Universit\`a degli Studi di Trieste,
via G.B. Tiepolo 11, 34143 Trieste, Italy
\and
Astrophysikalisches Institut Potsdam, An der Sternwarte 16, D-14482 Potsdam, Germany
\and
CRAL,UMR 5574: CNRS, Universit\'e de Lyon,
\'Ecole Normale Sup\'erieure de Lyon,
46 all\'ee d'Italie, F-69364 Lyon Cedex 7, France
\and
Kapteyn Astronomical Institute, Postbus 800, 9700 AV Groningen
\and
Center for Astrophysics and Space Astronomy, University of Colorado
389 UCB  (CASA),  Boulder, CO 80309-0389
 }

\mail{}
\authorrunning{Caffau et al. }
\titlerunning{The solar photospheric carbon abundance}

\date{Received ... / Accepted ...}
\abstract
{The use of hydrodynamical simulations, the selection of atomic data,
and the computation of deviations from local thermodynamical equilibrium
for the analysis of the solar spectra have implied a downward revision of the
solar metallicity. We are in the process of using
the latest simulations computed with the CO5BOLD code 
to reassess the solar chemical composition. Our previous 
analyses of the key elements oxygen and nitrogen have not
confirmed any extreme downward revision of $Z$.}
{We determine the solar photospheric carbon abundance by using a 
radiation-hydrodynamical CO5BOLD model,
and compute the departures from local thermodynamical equilibrium
by using the Kiel code.}
{We measure equivalent widths
of atomic \ion{C}{i} lines on high resolution,
high signal-to-noise ratio solar atlases of disc-centre intensity
and integrated disc flux.
These equivalent widths are analysed with the use of our
latest solar 3D hydrodynamical simulation computed
with CO5BOLD. Deviations from local thermodynamic
equilibrium are computed in 1D with the Kiel code, using
the average temperature structure of the hydrodynamical simulation
as a background model.}
{Our recommended value for the solar carbon abundance,
relies on 98 independent measurements of observed lines
and is A(C)=$8.50\pm 0.06$, the quoted error is 
the sum of statistical and systematic error. 
Combined with our recent results for
the solar oxygen and nitrogen abundances this implies 
a solar metallicity of $Z=0.0154$ and $Z/X=0.0211$.}
{Our analysis implies a solar carbon abundance which is
about 0.1\,dex higher than what was found in previous analysis
based on different 3D hydrodynamical computations.
The difference is partly driven by our equivalent width measurements
(we measure, on average, larger equivalent widths with respect to the 
other work based on a 3D model), in part it is likely due
to the different properties of the hydrodynamical simulations
and the spectrum synthesis code.
The solar metallicity we obtain from the CO5BOLD analyses is in slightly
better agreement with the constraints of helioseismology than the previous
3D abundance results.
}
\keywords{Sun: abundances -- Stars: abundances -- Hydrodynamics -- Line: formation}            

\maketitle{}           

\section{Introduction}

The importance of an accurate knowledge of the solar abundances can hardly
be overstated since they serve as the reference for all other celestial 
objects.
The high performance of the new-generation instruments
allows to derive accurate
stellar abundances and therefore 
the requested accuracy of the reference solar abundances is 
increased. This can, at least partly, 
explain the current revival in spectroscopic solar 
abundance studies. The very large gap in
resolution between solar and stellar
spectra, that existed until a few decades ago, 
is diminishing rapidly.
The majority 
of recent solar abundance determinations
relies on observational data that are almost 30
years old, both for the disc-centre intensity and for 
the integrated disc flux (Jungfraujoch grating spectra and 
Kitt Peak Fourier Transform Spectra, respectively).

For a long time, solar abundances 
were considered well established, and only 
minor refinements were suggested
by each new study, usually driven by improved
atomic or molecular data.
By using atomic or molecular lines, or both, the many analyses 
of the photospheric solar carbon made in 1980-2000, 
were converging toward the value of  
A(C)=8.52 $\pm 0.06$ \citep{grevesse98},
which was slightly lowering 
the previous values by including the appropriate NLTE
corrections.

However, \citet{allende02} announced an 
important downward revision of the C abundance from the analysis 
of the forbidden [CI] 872.7\pun{nm} line (A(C)$=8.39\pm 0.04$).
A subsequent paper by \citet{asplund05} obtained a similar
downward revision of the carbon abundance, also when using permitted 
atomic and molecular lines.

The new abundances of C, as well as those of other elements, 
are in conflict with some solar properties; 
solar models \citep{yang07}
and helioseimology \citep{basu,chaplin08,delahaye06} cannot be reconciled
with the recent revision of solar abundances by \citet{sunabboasp}. 

Solar abundances of the light elements, which have the highest cosmic 
abundance, are particularly important to understand stellar and galactic 
composition. Besides being the main contributors
to the solar metallicity $Z$, the CNO abundances are useful
to study the depletion in the interstellar gas (ISM).
For example, the comparison of the C/O ratio in the ISM
with that in the solar photosphere tells us how much
C has been locked into dust.
Also the study of diffusion effects in the corona and solar wind
requires the use of photospheric solar abundances
as a reference. Carbon, being a highly volatile element, partly escaped 
from carbonaceous chondrites,
so that  the solar system abundance of C relies mainly on the analysis of the
photospheric spectrum. 

The solar spectrum is rich in atomic C lines as well
as in lines of C-bearing molecules. Due to its high first
ionisation potential (11.26\pun{eV}), the measurable lines 
of carbon in the Sun are only those of \ion{C}{i}.  
Several tens of \ion{C}{i}
lines are present in the visual and near IR spectrum, but only few are suitable
for abundance analysis. The chosen lines should be weak, unblended, with
accurately known transition probabilities and, ideally, formed in LTE. 
Strong lines, with large equivalent width (EW$\geq$15-20\pun{pm}) should be 
rejected because the collisional damping constants are uncertain.
Only one forbidden line, at 872.7126\pun{nm}, has been detected in the 
solar spectrum.

Molecular lines are highly temperature sensitive 
and require a very accurate analysis of the photospheric thermal 
structure as the one made by \citet{ayres06} for the infra-red CO
features. \citet{hhoxy}
prefers to consider only atomic lines, while
\citet{grevesse87}
derive the abundances from the 
vibration-rotation and pure rotation lines of the CO and 
CN diatomic molecules to be more accurate.

In the present paper we analyse only 
\ion{C}{i} atomic transitions to derive the
solar photospheric carbon abundance.

\section{Selection of lines}
\label{s:selection}

As a starting point, we looked at a sample obtained by combining the
\ion{C}{i} lines 
chosen by \citet{grevesse91}, \citet{biemont93}, \citet{takeda94}, and
\citet{asplund05} (see Table~\ref{allcilines}).
We examined these lines, compared the available solar atlases among them and 
to synthetic profiles. We excluded from our analysis the lines that we judged 
too heavily blended (e.g.\ the line at 477.0\,nm) 
compared to the synthetic spectra, or the lines for which the
disagreement among observed data was unexplained and too large 
(e.g.\ the line at 1180.1\,nm). 
Furthermore, we eliminated the lines
whenever we suspected a significant contamination 
from telluric absorption, based on the comparison 
of the observed atlases and synthetic spectra, and
also on the inspection of spectra of fast rotating stars
indicating the presence of telluric lines (e.g. the line at 1602.1\,nm).
The excluded lines are flagged by ``3'' in the columns ``Quality'' of Table~\ref{allcilines}.
The final list of our sample of lines, labelled as Quality ``1'' or ``2'' in Table~\ref{allcilines},
is given in Table~\ref{ac}.
We labelled as ``1'' the lines that are not blended, or the blends are negligible
in comparison to the \ion{C}{i} line, or we think we are able to model the blends.
We labelled as ``2'' the lines we are less confident in.
These lines show differences in the observed spectra (e.g. the line at 711.1\,nm)
or we can hardly reproduce their shape with a synthetic profile (e.g. the line at 1734.6\,nm)
or we are not confident to be able to take into account the telluric absorptions
(e.g. the line at 1778.9\,nm).
The final selection consists of 45 individual lines for which 
we have 98 EW measurements. 
The subsample of good data, labelled ``Quality=1'',
contains 25 lines, 66 measured EWs.

\begin{table}
\caption{Lines considered for the abundance determination.}
\label{allcilines}
\begin{center}
\begin{tabular}{rcrc}
\noalign{\smallskip}\hline\noalign{\smallskip}
$\lambda$ & Quality & $\lambda$ & Quality\\
nm & & nm & \\
\noalign{\smallskip}\hline\noalign{\smallskip}
  477.000~~   & 3 & 1174.822     & 2 \\
  477.5907    & 2 & 1177.754     & 1 \\
  505.2167    & 1 & 1180.110     & 3 \\
  538.0336    & 1 & 1184.873     & 1 \\
  658.7608    & 2 & 1186.299     & 1 \\
  708.5511    & 3 & 1189.291     & 1 \\
  708.7827    & 2 & 1189.575     & 1 \\
  711.1475    & 2 & 1254.948     & 2 \\
  711.3180    & 1 & 1256.212     & 1 \\
  713.2112    & 3 & 1256.904     & 1 \\
  783.7105    & 3 & 1258.159     & 1 \\
  801.8564    & 1 & 1261.410     & 3 \\
  833.5149    & 1 & 1602.164     & 3 \\
  872.7126    & 2 & 1704.516     & 3 \\
  875.3079    & 3 & 1723.448     & 3 \\
  887.3390    & 3 & 1734.638     & 2 \\
  906.1432    & 1 & 1744.860     & 1 \\
  907.8278    & 1 & 1745.597     & 1 \\
  911.1797    & 1 & 1747.591     & 3 \\
  918.2831    & 3 & 1750.564     & 1 \\
  960.3032    & 2 & 1755.446     & 3 \\
  962.0795    & 3 & 1763.738     & 2 \\
  965.8435    & 1 & 1778.960     & 2 \\
 1012.3871    & 1 & 2102.313     & 2 \\
 1068.5345    & 1 & 2121.155     & 2 \\
 1070.7333    & 1 & 2125.989     & 2 \\
 1072.9533    & 1 & 2290.656     & 2 \\
 1075.3985    & 1 & 3085.462     & 3 \\
 1161.929~~   & 3 & 3129.748     & 2 \\
 1163.050~~   & 3 & 3406.579     & 3 \\
 1165.884~~   & 3 & 3991.177     & 3 \\
 1165.968~~   & 3 &\\           
\noalign{\smallskip}\hline\noalign{\smallskip}
\end{tabular}
\\
\end{center}
Quality: 1 good line, 2 line with problems, 3 line rejected
\end{table}

For the abundance determination one could rely on line profile fitting
or on EW measurements. 
The line profile fitting procedure has many advantages due to the fact
that not only the strength of the line is taken into account, but
also the line shape. In case the synthetic line profile provides
a faithful reproduction of the line shape, we consider this procedure 
superior.
We stress here that by
``line profile'' fitting we mean fitting with a synthetic profile, computed
using all the known lines in the range. But ``fitting'' with a synthetic
profile consisting of a single line is conceptually identical
to measuring the EW by fitting with a Gaussian or Voigt profile, 
although it has the advantage of treating correctly the line asymmetry
which, however, is in general irrelevant for abundance work.
If poorly known blends interfere, the EW measurement procedure with 
deblending (see below) is the more secure option.
The present analysis is based on EW measurements. We give preference to this 
approach due to the following problems with the 
available \ion{C}{i} lines:
\begin{itemize}
\item a large fraction of the lines are blended, and the atomic data 
of the contaminants are not well known, so that when included in the
3D synthetic spectra, the comparison with the observed spectra
is not reliable; the measurement of the EW, on the other hand,
can be reliable, since the extra absorption can be modelled by a suitable
gaussian or Voigt profile;
\item some of the lines are contaminated by telluric absorption; also
in this case, the contaminating telluric absorption can be modelled
as above providing a reliable measure of the EW;
\item for some lines the continuum placement is problematic, due
to the presence of neighbouring lines, whose atomic data are often
poorly known; the EW measurement with {\tt splot}, on the other hand, 
is designed to handle such situations;
\item NLTE effects, not taken into account in the 3D synthetic profile, 
can change the shape of the line.
\end{itemize}
The last point has not yet been investigated in detail since no
3D-NLTE analysis for carbon is available at the moment.
But we expect that carbon does not behave differently from oxygen, and
it is shown in \citet{asplund04} that the 3D-NLTE line profile
is different from the 3D-LTE one.
One could consider to use the line profile fitting technique for clean lines,
that form a small subsample of the complete set of lines,
reserving the EW measurement to the ``problematic'' lines.
However, in this way the analysis would not be homogeneous 
over the complete sample of lines. The adopted method of
EW measurements also allows a more direct comparison with other 
analyses available in the literature.

For the measurement of the EW we used the IRAF\footnote{IRAF is 
distributed by the National Optical Astronomy Observatories,
which are operated by the Association of Universities for Research
in Astronomy, Inc., under cooperative agreement with the National
Science Foundation.}\citep{tody} task 
{\tt splot}. In the case of blended lines
we used the deblending option of {\tt splot},
that permits to fit the spectral profile with a number of
Gaussian and/or Voigt functions. In this way any known line in a range
can be simulated with a theoretical profile.
Generally for weak lines we used a Gaussian profile
to fit the observed profile, while for
strong lines we used a Voigt function.
For unblended lines we also used direct integration.
We are aware that the observed profile is asymmetric
while both Gaussian and Voigt functions are symmetric.
Several experiments convinced us
that the use of a Voigt profile to measure the EW of an asymmetric
3D profile differs from the real EW by less than 1\,\%.
This error is surely negligible when compared to the uncertainty
of the EW measurement due to the continuum placement, which in the 
case of a typical  observed spectrum can be up to 5\,\%.

For the majority of the lines the equivalent widths we obtain are close to 
the values of \citet{biemont93}, but  not for all.
We could compare only the Delbouille disc-centre spectrum, which is
the observed data considered in \citet{biemont93}.
In principle, strong lines should be rejected because of uncertain values 
of NLTE corrections and line broadening parameters which become important.
We keep these strong lines anyway in the sample,
because they do not disagree with the other lines, and there is no evident
trend of the abundance as a function of the equivalent width.

When available, we used \loggf\ from NIST \citep{wiese96}, as retrieved 
from the ASD database \citep{ralchenko}. 
The values are given in Table~\ref{ac}.
All \loggf\,-values used in \citet{biemont93}, except the one of the 
801.8\pun{nm} line, are very close to the values of NIST.
For our sample of \ion{C}{i} lines the NIST database relies on four sources
\citep{luo89,hibbert93,nussbaumer84,weiss96}, the main one being 
\citet{hibbert93} which covers all the lines.

For the Van der Waals broadening constants we proceeded as in \citet{oxy}. 
When available (for 35 lines of our sample) we rely on \citet{abo4} values.
For the remaining lines we used the WIDTH approximation,
implemented in the Kurucz routine WIDTH \citep[see][]{ryan}.
If we remove the lines without Van der Waals broadening constants
from \citet{abo4}, the derived carbon abundance is 
less than 0.03\,dex higher than when considering the 
complete sample of lines. Therefore we decided to keep all the 
lines we selected for the abundance determination.


\section {Equivalent widths in the literature}

Reliable observed EWs are those measured by \citet{grevesse91}
and by \citet{biemont93} on the disc-centre Jungfraujoch Atlas.
The values used by \citet{stuerenburg} are taken from the EW 
measurements by \citet{baschek67} which were based on old atlases, 
and the ones in \citet{asplund05} are the EWs of the synthetic 
best fit profile.
Only two lines have been considered in all four analyses, 
and they are presented in Table\,\ref{ewlit}.
 
\begin{table}
\caption{Comparison of the disc-centre EW of 
two \ion{C}{i} lines as determined by different authors.
}
\label{ewlit}
\begin{center}
\begin{tabular}{rcclrrr}
\noalign{\smallskip}\hline\noalign{\smallskip}
$\lambda$ (nm) & \multicolumn{3}{c}{EW (pm)}\\
               & G91 & B93 & A05 & C10\\
\noalign{\smallskip}\hline\noalign{\smallskip}
 960.30  & 10.8 & 9.62 & 9.6 & 11.5\\
 2102.31 & 10.0 & 10.26 & 8.76 & 10.0\\
\noalign{\smallskip}\hline\noalign{\smallskip}
\end{tabular}
\end{center}
Notes: G91: \citet{grevesse91}, B93: \citet{biemont93}, 
A05: \citet{asplund05}, and C10: this work.
\end{table}

For the line at 960.3036\pun{nm}, we find a significantly larger EW than
the other three authors, whose results agree closely. On the other hand,
our EW for the line at 2102.3151\pun{nm} is very similar to those by
\citet{grevesse91} and \citet{biemont93}, while the theoretical EW derived 
by  \citet{asplund05} from their 3D model by using their best fit abundance 
is much lower (see Table~\ref{ewlit}), even though the measured EWs should 
be corrected for blending.

We note that \citet{biemont93} gave a low weight to both of these lines,
presumably because they are affected by telluric absorption and
other blends, and hence reliable equivalent widths are difficult to measure.
Nevertheless, Table~\ref{ewlit} demonstrates once again that equivalent width
measurements differ considerably from author to author and are a major source 
of uncertainty.

All the investigations of the solar carbon abundance cited above rely on a 
single solar atlas. In fact, as already mentioned in \citet{oxy}, the 
available solar atlases do not always agree. This could be due to telluric 
absorption, to variability in solar spectrum, or to systematic effects related
to the different observations. The present analysis is based on four 
different solar spectra and hence should yield more reliable abundances.

\section{Observed spectra}

We considered the same four observed solar atlases publicly available
that we already used in \citet{oxy}. For disc-centre, this is the
double-pass grating spectrum taken at Jungfraujoch by \citet{delbouille}, 
ranging from 300 to 1000~nm, and the infrared FTS spectrum taken at 
Kitt Peak by \citet{delbouilleir}, covering the wavelength range 1000 
to 5400~nm (together called Delbouille intensity, {\bf DI}). In addition, 
the disc-centre FTS spectrum published by \citet{neckelobs} is used 
(330 to 1250~nm, Neckel intensity, {\bf NI}). \citet{neckelobs,neckel1999} 
also provide a corresponding FTS spectrum for the integrated disc flux 
(Neckel flux, {\bf NF}). Another set of Kitt Peak FTS scans by Brault and 
Testerman has been made available by \citet{kuruczflux} (300 to 1000~nm, 
hereafter Kurucz flux, {\bf KF}). For one line we resorted to the ATMOS 
solar atlas \citep{atmos,farmer}.

\section{Model atmospheres}

We used the same radiation-hydrodynamical model,
computed with the \cobold\ code \citep{freytag02,freytag03,Wedemeyer} 
used in our previous solar abundance analyses.
Details of the model can be found in \citet{zolfito,oxy}.
The same holds for the employed 1D models.
As a reference model we used a plane parallel 1D model 
(\xx\, with mixing-length parameter of 1.0) that shares the
micro-physics and radiation transfer scheme with \cobold.
We also used a 1D model obtained by temporal and horizontal 
averaging of the 3D hydrodynamic structure on surfaces of 
equal optical depth (\mD), as well as the Holweger-M\"uller
semi-empirical model \citep{hhsunmod, hmsunmod}.
When necessary this was put on a column mass scale,
assuming the same abundances and opacities as in the \cobold\ model.
For the spectrum synthesis based on the 1D models, we adopted a 
micro-turbulence of 1.0\,km s$^{-1}$, both for disc-centre (intensity) 
and integrated disc (flux) spectra.

\section{Results}
\subsection{The [CI] line at 872.7\, nm}

There is only one observable forbidden [CI] line, located at 872.7126\pun{nm} 
(2p$^2$ $^1$D$_2$ - 2p$^2$ $^1$S$_0$), with a lower level excitation potential 
of 1.264\pun{eV}. This line is important, being weak and therefore insensitive 
to the assumption about the damping constant, and according to \citet{stuerenburg} 
not affected by NLTE.

We measured the EW on the two disc-centre and two integrated disc solar spectra.
The result is EW(DI)=0.511\pun{pm}, EW(NI)=0.509\pun{pm}, EW(NF)=0.517\pun{pm}, 
EW(KF)=0.508 \pun{pm}.
We subtracted the contribution of the \ion{Fe}{i} blending line
($\lambda$=872.7132\pun{nm}, \loggf =-3.93, $\chi$ =4.186\pun{eV}, EW(Int)=
0.040\pun{pm}, EW(Flux)=0.045\pun{pm}) according to the 3D computation.
The [CI] line is formed in LTE, so that the LTE abundance derived from
the four observed solar atlases should be in close agreement. While the two 
disc-centre and the two integrated disc spectra are in very good agreement 
with each other, we find a difference in the carbon abundance
of 0.04\pun{dex} between disc-centre and integrated disc spectra. 
This effect could be attributed to NLTE effects on the
blending iron line which can be different for disc-centre and integrated disc 
spectra.

\subsection{Permitted lines}

The abundances derived from all the lines of our sample
have been assembled in Table \ref{ac}. Atomic data
for each line, the measured EWs, and the derived LTE abundances
using the 3D, \mD, and \xx\ models are provided in the table.
The total 3D correction, defined as 3D -\xx\, is positive for all 
the lines, except for the 477.5907\,nm line.
The so-called granulation correction, quantified by the difference
3D-\mD\, measures the effect of the horizontal temperature fluctuations.
It is never very large, and may be in the same direction 
as the total 3D correction, or in the opposite direction, depending on 
the line.
All our permitted \ion{C}{i} lines originate from highly excited lower levels 
($\chi > 7$\pun{eV}).
The 3D-\mD\ corrections for the weaker lines (EW $<4.0$\pun{pm}) 
in the optical and near infra-red range (450\pun{nm}$<\lambda <$820\pun{nm})
are small and mostly negative (3D-\mD $<0.01$).
We can compare these 3D-\mD\ corrections to the value $\Delta_{\rm gran}$
defined in \citet{mst02}.
The hydro-simulation they consider is a 2D model, while the 1D model is 
the temporal and horizontal average over surfaces of equal optical depth
of their 2D model, meaning that it is similar to our \mD\ model.
The \ion{C}{i} lines investigated in \citet{mst02} 
have $\lambda$=550\pun{nm} and EW$\approx $ 0.5\pun{pm}.
For $\chi > 6$\,eV, $\Delta_{\rm gran}$ is indeed slightly negative,
(see their Table\,1 and their Figure\,5) in qualitative agreement with 
the results found for comparable \ion{C}{i} lines in the present study.

Both 3D corrections (3D -\xx\ and 3D-\mD) increase with EW,
possibly indicating an inadequate choice of the micro-turbulence
parameter for the 1D models. We cannot discern any trend with 
the excitation energy or the wavelength.

We do not yet have the capability of computing
the deviations from local thermodynamic equilibrium (NLTE effects) 
in the 3D spectrum synthesis,
and, to our knowledge, such calculations have not yet been performed elsewhere.
As a first approximation we have therefore computed 1D NLTE corrections
using the \mD\ model as a background model. 
For each line we computed NLTE corrections with the 
Kiel code \citep{SH} and the model-atom of \citet{stuerenburg}.
The line blanketing is treated with an opacity distribution function
as provided by \citet{ck03}, assuming solar metallicity and a micro-turbulence 
of 2\,\kms.
We considered three possible choices for the parameter ${\rm S_H}$ 
quantifying the thermalizing effect of collisions with neutral hydrogen
according to the generalised Drawin approximation \citep{Drawin} 
as proposed by \citet{SH}:
\begin{enumerate}
\item classical scaling (${\rm S_H}=1$);
\item no effect of collisions with neutral H (${\rm S_H}=0$);
\item intermediate collisional efficiency (${\rm S_H}=1/3$).
\end{enumerate}

The NLTE correction obtained for each line is listed in Table\,\ref{ac}.
There is a general good agreement with the NLTE computations
of \citet{stuerenburg} and \citet{asplund05}, with a maximum 
difference of 0.02\,dex. We recall that these studies rely on a
different model atmosphere for the NLTE computation, so that
differences of a few hundredth of dex can easily be attributed to the
different input solar model.
This agreement in the 1D-NLTE computations is encouraging. As long as
no 3D-NLTE computation is available, it is certainly justified to apply this
1D-NLTE correction to our 3D-LTE abundances, as also done by \citet{asplund05}
in their careful work.

\begin{figure*}
\resizebox{\hsize}{!}{\includegraphics[clip=true,angle=0]{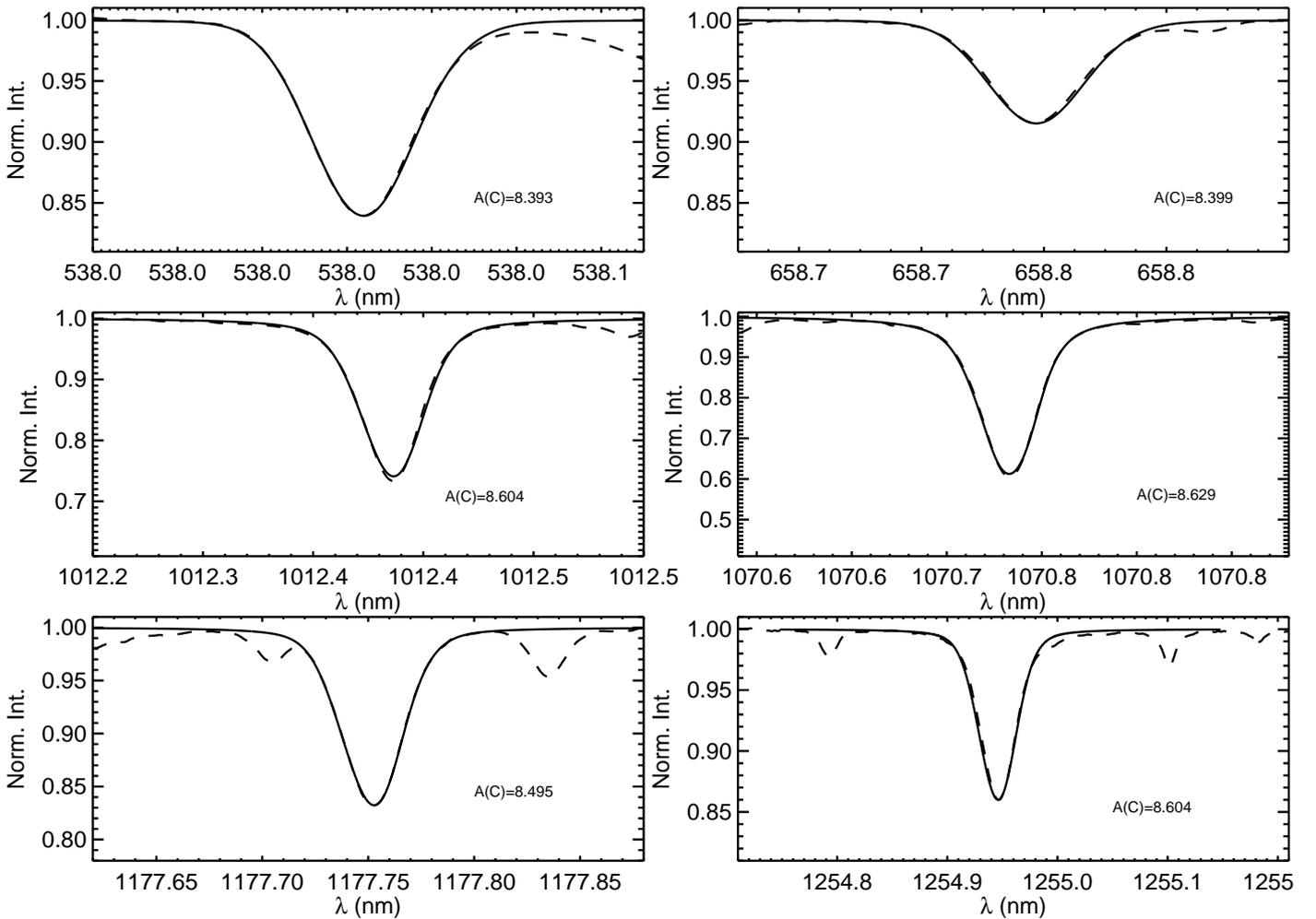}}
\caption{Comparison of the observed disc-centre solar spectrum
(dashed lines, DI for the two reddest lines and NI for the others)
with the corresponding 3D synthetic profiles (solid lines) for a selection of
clean \ion{C}{i} lines.
}
\label{comp3dobs}
\end{figure*}

\begin{figure*}
\resizebox{\hsize}{!}{\includegraphics[clip=true,angle=0]{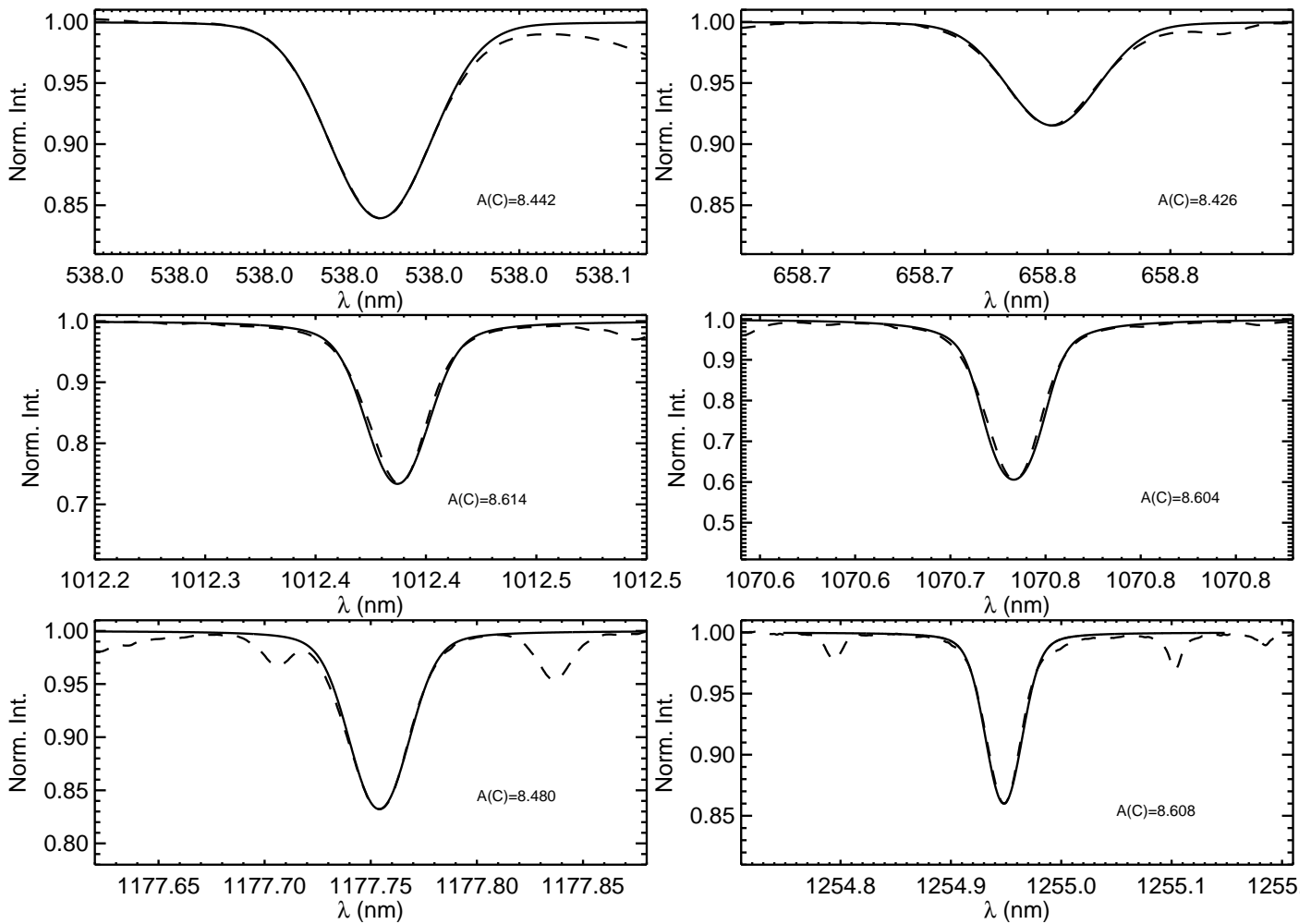}}
\caption{As Fig.\,\ref{comp3dobs}, but showing the HM synthetic profile (solid lines) 
superimposed on the observed solar spectrum (dashed lines).
To be consistent with Fig.\,\ref{comp3dobs}, the synthetic profile
is computed with \linfor, ignoring any blending lines.
}
\label{comphmobs}
\end{figure*}

As explained above, we rely on EW measurements for the carbon abundance
determination, because a considerable fraction of the lines is blended,
and NLTE effects are non-negligible. Nevertheless, it is useful to compare
the observed spectrum with the 3D synthetic profiles for some of the cleanest 
lines. A few examples are shown in Fig.\,\ref{comp3dobs}.
The agreement is encouraging, but the abundance needed to achieve the best (visual)
agreement between 3D-synthetic and observed line profile is not always identical
to the abundance obtained from matching the EWs.
This can be due to remaining blends, not included in the 3D-synthetic profile, NLTE
effects that can change the shape of the line profile, or the adopted damping constants.
Comparison of Figs.\,\ref{comp3dobs} and \ref{comphmobs} shows that the
3D synthetic profiles generally can reproduce the observed profiles of the selected clean
\ion{C}{i} lines somewhat better than the 1D  synthetic profiles calculated from the 
HM model with a micro-turbulence of 1.0\,\kms.

\subsection{The solar carbon abundance}

The final carbon abundance depends
only weakly on the assumption made about ${\rm S_H}$.
Applying the NLTE corrections to the 3D LTE results, and computing
the average of the abundances of Table\,\ref{ac}, we obtain:

\begin{equation}
\begin{array}{c @{\rm A(C)=~} c @{~~~~~{\rm for} ~~~~~ {\rm S_H} =} c }
     & 8.446\pm 0.121 & 0\\
     & 8.498\pm 0.110 & 1/3\\
     & 8.523\pm 0.112 & 1\\
\end{array}
\end{equation}
For reference, the results for the HM model are:

\begin{equation}
\begin{array}{c @{\rm A(C)=~} c @{~~~~~{\rm for} ~~~~~ {\rm S_H} =} c }
     & 8.449\pm 0.135 & 0\\
     & 8.503\pm 0.116 & 1/3\\
     & 8.532\pm 0.112 & 1\\
\end{array}
\end{equation}

\begin{figure}
\resizebox{\hsize}{!}{\includegraphics[clip=true,angle=0]{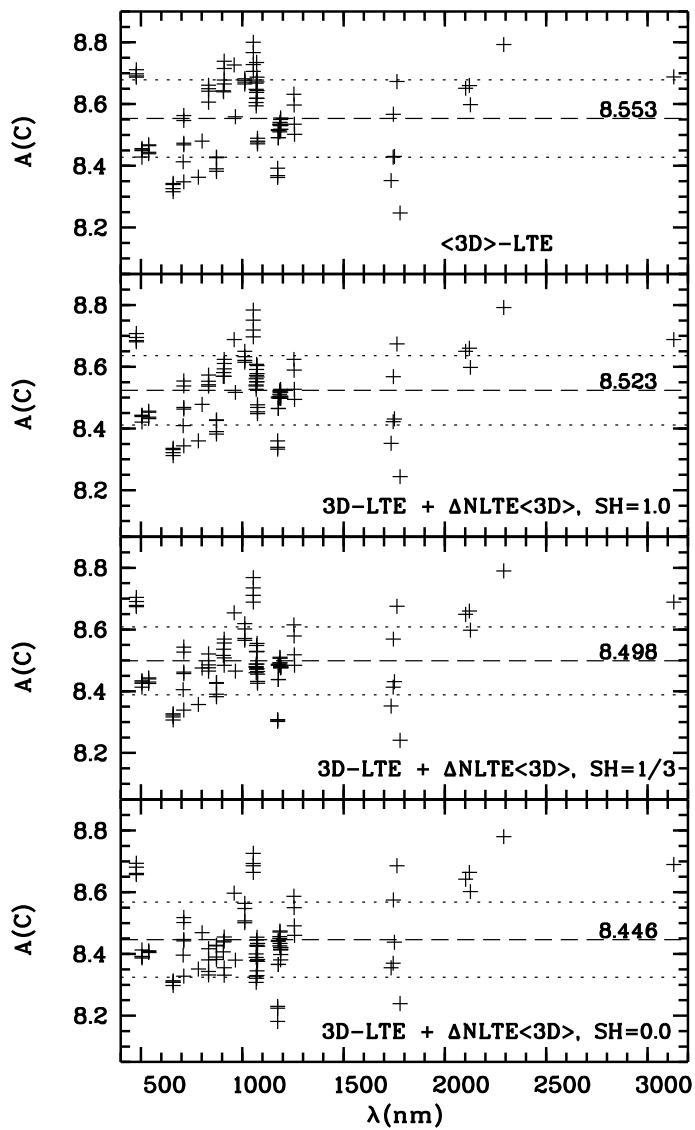}}
\caption{The carbon abundance as a function of wavelength of the individual 
\ion{C}{i} lines compiled in Table~\ref{ac}, as obtained from the 3D solar 
model for different assumptions about the 1D NLTE corrections, which are
all based on the \mD\ model.
}
\label{ac3d}
\end{figure}

\begin{figure}
\resizebox{\hsize}{!}{\includegraphics[clip=true,angle=0]{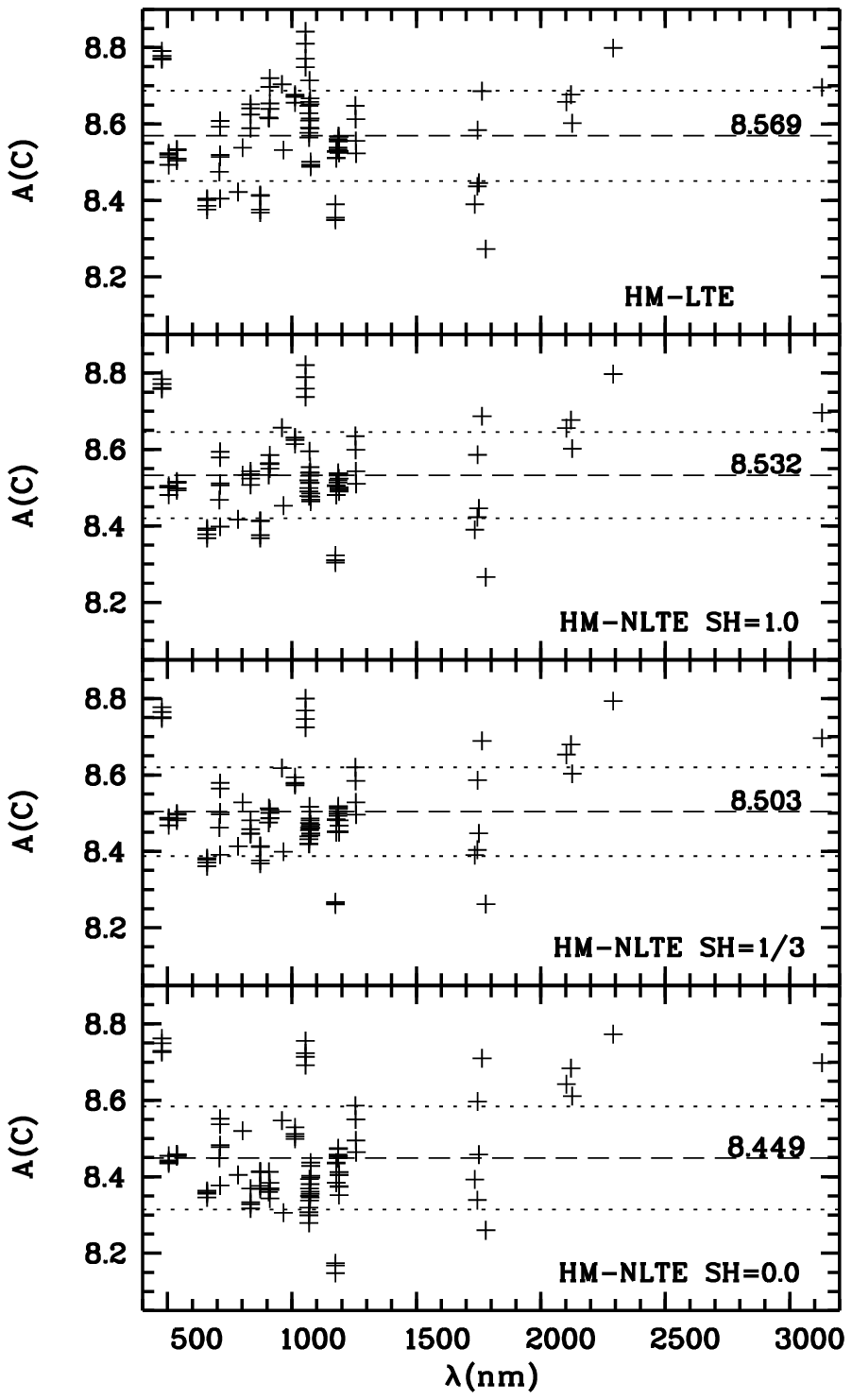}}
\caption{As Fig.\,\ref{ac3d}, but showing the carbon abundances obtained from
the HM model.
}
\label{achm}
\end{figure}

The results from \cobold\ and HM models are in very good agreement.
The carbon abundances from the various lines
as a function of wavelength  
are shown in Fig.~\ref{ac3d} for the \cobold\ model,
and in Fig.~\ref{achm} for the HM model.

Our favoured value is A(C)$=8.498\pm 0.110$, 
obtained applying the NLTE correction, with ${\rm S_H} =1/3$,
to the 3D-LTE abundance.
If we restrict the abundance determination to the lines labelled as 
``Quality=1'' in Table\,\ref{allcilines}, the result is 
A(C)$=8.490\pm 0.048$. 
The carbon abundance from the subsample is very close to the one
obtained from the complete sample, only 0.008\,dex smaller, 
while the line-to-line scatter is much reduced.
Nevertheless, we prefer the result from the complete sample,
being aware that the selection of ``good'' lines is somehow 
subjective, and hardly changes the mean carbon abundance.

\section{Discussion}

If we consider the EWs of 55 lines in \citet{biemont93}, based on Delbouille disc-centre 
spectra, together with the \loggf\ -values used in that work, we find 
A(C)=$8.518\pm 0.137$ from our 3D model. If we use instead the updated \loggf\ -values
from the NIST database, we obtain  A(C)=$8.504\pm 0.125$. 
This latter value can be compared to our LTE result, based on the 40 lines in common 
with \citet{biemont93} and measured in the same solar atlas, 
of A(C)=$8.535\pm 0.121$ obtained with the 3D model,
and of A(C)=$8.550\pm 0.108$ obtained with the HM model.
The LTE abundance based on our complete sample of 98 lines is
A(C)=$8.553\pm 0.125$ from the 3D model and A(C)=$8.569\pm 0.118$ from the HM model.
Our abundance is slightly higher than the one of \citet{biemont93} because of
the line selection and differences in the EWs for some lines, but the overall agreement
is very satisfactory.

3D-NLTE abundances are obtained by applying the 1D-NLTE corrections with 
${\rm S_H}=1/3$ to the individual 3D-LTE abundances. 
All 98 3D-NLTE abundances lie within 3$\sigma$ of the mean value,
A(C)$=8.498\pm 0.110$. Keeping only abundances within 2$\sigma$ of the mean value,
94 values meet the cutoff, and the average becomes A(C)$=8.492\pm 0.098$; 
within 1$\sigma$, still 70 values survive, and A(C)$=8.485\pm 0.049$. As expected,
the standard deviation becomes smaller, but the average remains almost the same.

To check the validity of the EW approach, we used line profile fitting 
to determine the carbon abundance for two lines which are not blended, 
are weak, and have a very small NLTE correction: 538.0\pun{nm} and 658.7\pun{nm}. 
On average, the result is about 0.02\pun{dex} below the one obtained from the EW.
For the majority of the selected lines the atomic data of the blending components
are not very well known. These blending components are, however, sufficiently 
separated that EWs can be measured by fitting multiple Gaussian or Voigt profiles.
We therefore prefer to use EWs, rather than line profile fitting 
(see Sect.\,\ref{s:selection}). 

We find no obvious trend of the abundance with the lower level energy,
neither for the LTE nor for the NLTE results. However, this is not
surprising since all carbon lines of our sample originate from similar 
high excitation levels, the range in energy being little more than 2\pun{eV}.

\begin{figure}
\resizebox{\hsize}{!}{\includegraphics[clip=true,angle=0]{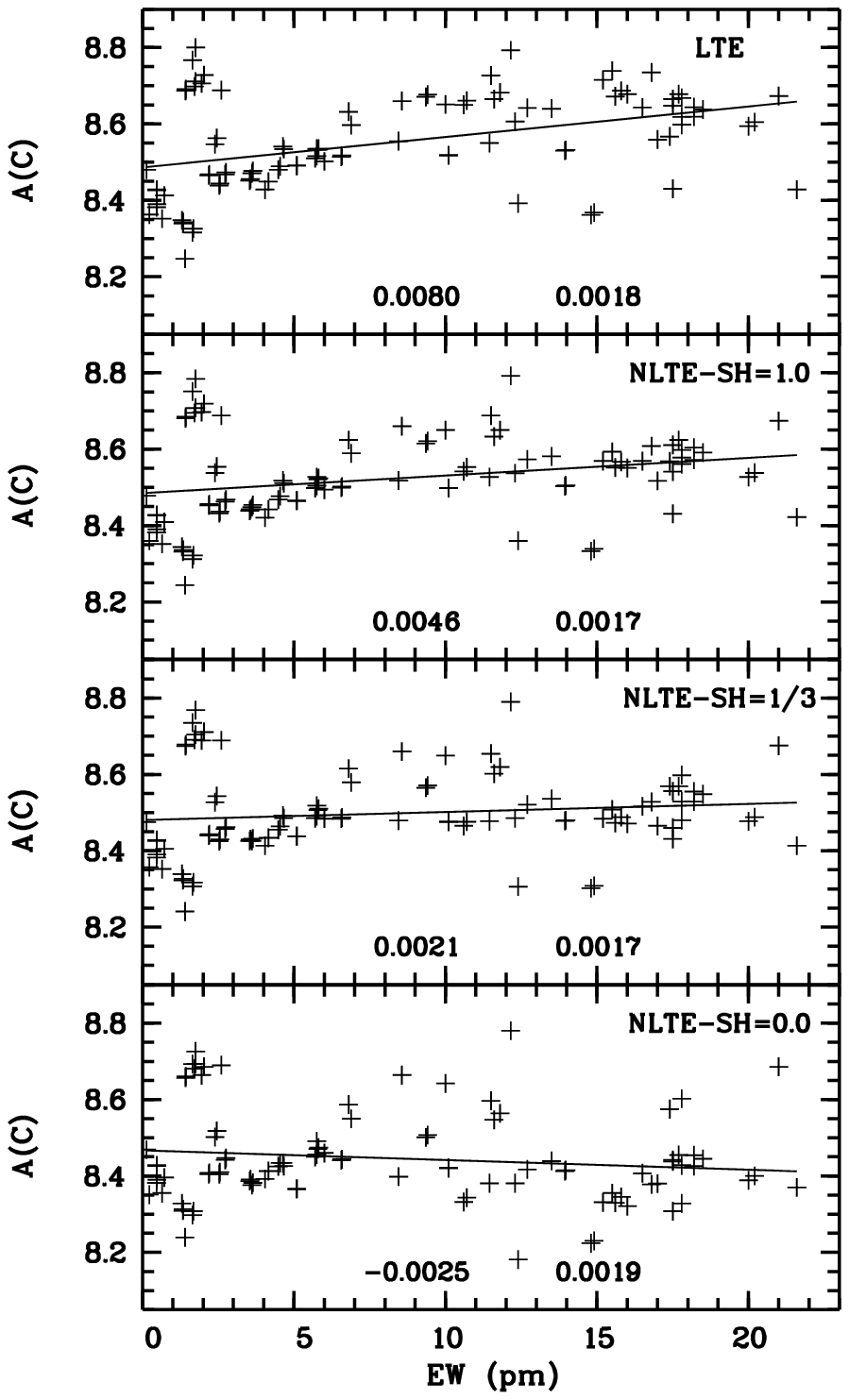}}
\caption{The carbon abundances as a function of EW for the 3D model.
The two number in the lower part of the plots indicate the slope of the 
best fit linear relation and its 1~$\sigma$ uncertainty.
}
\label{acew3d}
\end{figure}

\begin{figure}
\resizebox{\hsize}{!}{\includegraphics[clip=true,angle=0]{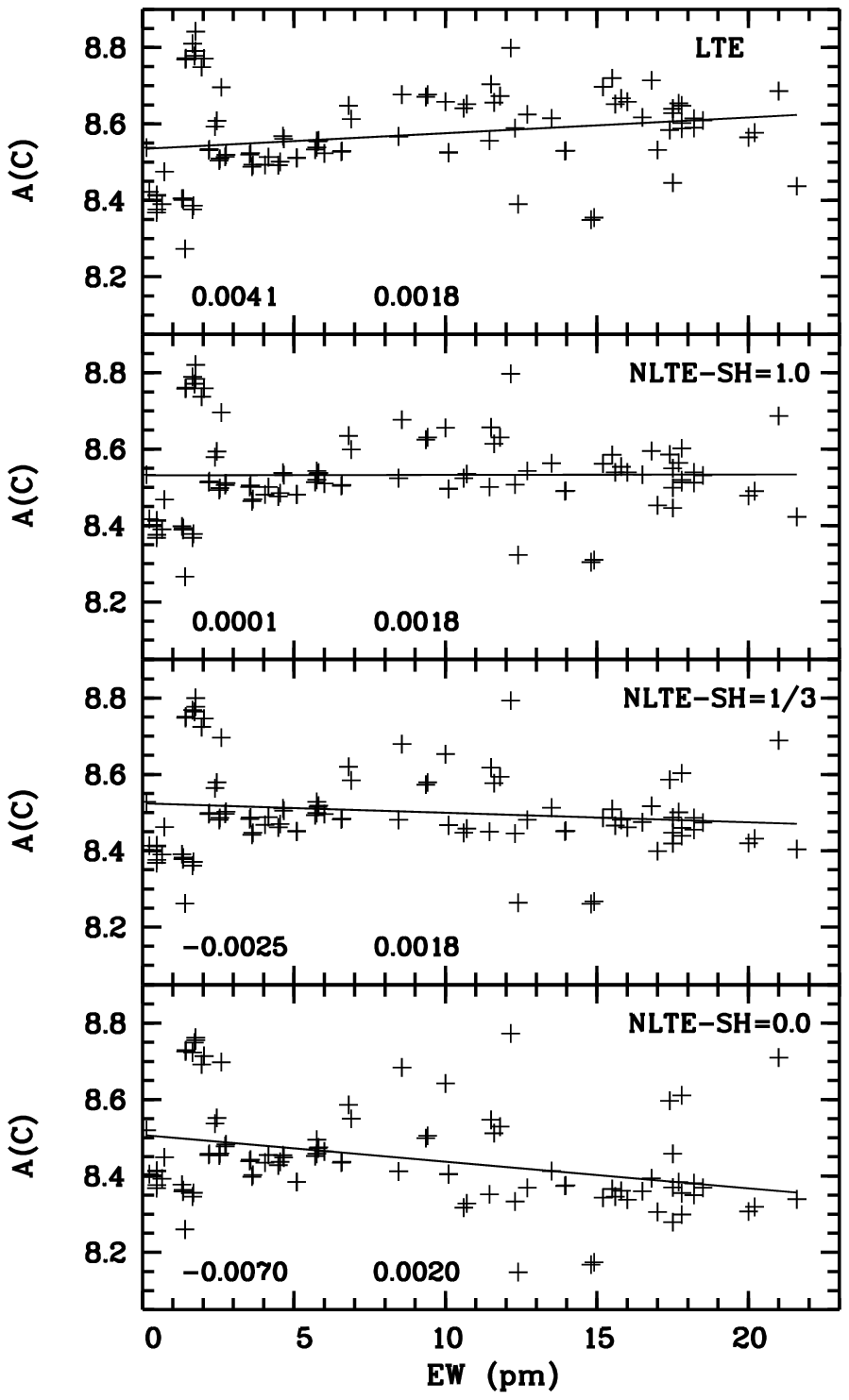}}
\caption{The carbon abundances as a function of EW for the HM model.
}
\label{acewhm}
\end{figure}

There is, however, a clear trend that the 3D-LTE abundances increase with 
equivalent width (see Fig.\,\ref{acew3d}). The trend in reduced or even reversed 
after application of the NLTE corrections, depending on the choice of  
${\rm S_{\rm H}}$. As illustrated in Fig.\,\ref{acew3d}, there is a slight 
negative trend of the 3D-NLTE abundance with EW for ${\rm S_{\rm H}}=0.0$, and a 
slightly positive one for ${\rm S_{\rm H}}=1/3$. The trend vanishes for a value 
of ${\rm S_{\rm H}}$ somewhere in the range $[0,1/3]$.
The corresponding results obtained using the HM model are shown in Fig.\,\ref{acewhm}.
The behaviour is similar to what is found with the 3D model, but the slope of
the A(C)--EW relations is systematically reduced (more negative) in all cases.
The results form the HM model suggest that the slope vanished for ${\rm S_{\rm H}}$
close to $1$. We note that the correlation A(C)$_{\rm LTE}$--EW persists even if we 
consider only weak lines, indicating that the slope is in fact due to NLTE effects.

Without available experimental data on cross sections for collisions with neutral hydrogen, 
one might be tempted to fix the value of ${\rm S_{\rm H}}$ empirically by requiring a
vanishing  slope in the A(C)--EW plane. Our 3D results would then suggest that 
$0.0<{\rm S_{\rm H}}<1/3$. However, we prefer to delay this conclusion until a 
complete 3D-NLTE computation will be available.
In the meantime, obliged to take a decision, we adopt the intermediate
value of ${\rm S_{\rm H}}=1/3$, which is the favourite value of the Holweger school.

\begin{figure}
\resizebox{\hsize}{!}{\includegraphics[clip=true,angle=0]{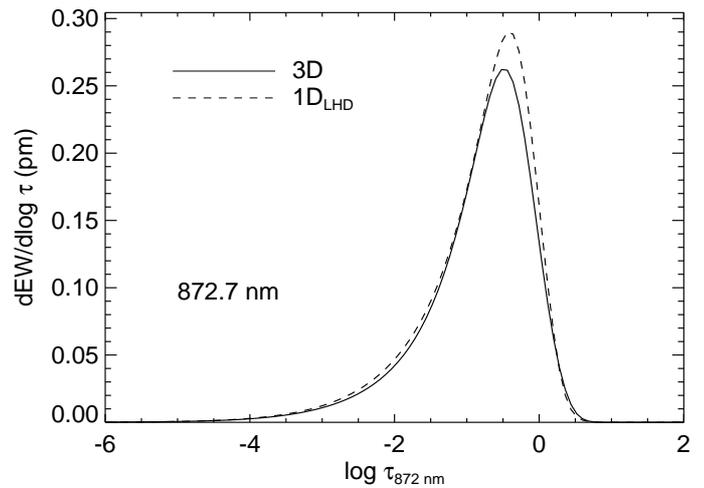}}
\caption{3D and \xx\ equivalent width contribution function at disc-centre 
for the [CI] line at 872.7~nm.
}
\label{contf}
\end{figure}

Our carbon abundances are larger than 
those derived by \citet{allende02} and
\citet{asplund05}. The difference is
striking if we consider the [CI] line. 
While the measured EWs are similar,
we adopt an EW which is about
12\% smaller than that of \citet{asplund05},
due to the correction for the blending
\ion{Fe}{i} line. 
In spite of this, our derived A(C) is 0.11\,dex higher
than that of \citet{asplund05}.
This must be ascribed to the difference between
the \cobold\ solar model + \linfor\ and the hydrodynamical
simulation and spectral-synthesis code employed by  \citet{asplund05}.
The difference between the models (see Figure\,1 in \citet{oxy}) 
affects both the mean
temperature structure and the temperature
fluctuations in the region where the [CI] line is formed.
Fig.\,\ref{contf} shows that the main contribution to the [CI] 
line absorption comes from the layers between $\log\tau = 0$, to $-2$
where the differences between two hydrodynamical simulations are largest.
Since the mean temperature gradient is steeper in the 3D model of
\citet{asplund05}, a lower carbon abundance is needed to reproduce the 
observed equivalent width, in qualitative agreement with the results
mentioned above.
The same behaviour has been noticed for the [OI] line at 636.3\,nm line 
(see \citealt{oxy} and \citealt{asplund04}).

The situation is similar
if we compare the average carbon abundance. 
From the permitted \ion{C}{i} lines, \citet{asplund05}
derive  $\langle {\rm A(C)}\rangle = 8.36 \pm 0.03$. 
This value must be compared with our value of $8.446\pm 0.121$,
which correspond to ${\rm S_H}=0$, consistent with the
assumption of \citet{asplund05}. Part of the difference is due 
to our EWs, which are generally close to those of \citet{biemont93}, 
and therefore larger than those of \citet{asplund05}. Part of the
difference is, however, due to the different hydrodynamical
model and/or the different spectrum synthesis code.

{\scriptsize
\begin{longtable}{rccrrlrrrrrrrr}
\caption{Line parameters, 3D and 1D carbon abundance (LTE) and 1D NLTE
corrections for our sample of selected \ion{C}{i} lines.}\\
\hline\noalign{\smallskip}
\hline\noalign{\smallskip}
\label{ac}
$\lambda$ & SP & $\chi$ & EW & \loggf & Acc. & \multicolumn{3}{c}{A(C) (LTE)} & \multicolumn{2}{c}{3D corrections} &  \multicolumn{3}{c}{1D-NLTE corrections}\\
nm & & eV & pm &  & & 3D & \mD\ & \xx\ & 3D-\mD\ & 3D-\xx & 1.0 & 1/3 & 0.0\\
\noalign{\smallskip}\hline\noalign{\smallskip}
\hline\noalign{\smallskip}
\endfirsthead
\caption{continued.}\\
\hline\noalign{\smallskip}
\hline\noalign{\smallskip}
$\lambda$ & SP & $\chi$ & EW & \loggf & Acc. & \multicolumn{3}{c}{A(C) (LTE)} & \multicolumn{2}{c}{3D corrections} &  \multicolumn{3}{c}{1D-NLTE corrections}\\
nm & & eV & pm &   & & 3D & \mD\ & \xx\ & 3D-\mD\ & 3D-\xx & 1.0 & 1/3 & 0.0\\
\hline\noalign{\smallskip}
\endhead
\hline\noalign{\smallskip}
\endfoot
\endlastfoot
  872.7126 & DI & 1.26 & 0.47 & $-8.140$      &B &  8.428 & 8.401 & 8.388   & $ 0.027$ & $ 0.040$ & $ 0.000$ & $ 0.000$ & $ 0.000$ \\
  872.7126 & NI & 1.26 & 0.47 & $-8.140$      &B &  8.426 & 8.399 & 8.386   & $ 0.027$ & $ 0.040$ & $ 0.000$ & $ 0.000$ & $ 0.000$ \\
  872.7126 & KF & 1.26 & 0.46 & $-8.140$      &B &  8.382 & 8.360 & 8.354   & $ 0.021$ & $ 0.028$ & $ 0.000$ & $ 0.000$ & $ 0.000$ \\
  872.7126 & NF & 1.26 & 0.47 & $-8.140$      &B &  8.390 & 8.369 & 8.362   & $ 0.021$ & $ 0.028$ & $ 0.000$ & $ 0.000$ & $ 0.000$ \\
  477.5907 & DI & 7.49 & 1.71 & $-2.304$      &C &  8.698 & 8.709 & 8.709   & $-0.011$ & $-0.011$ & $-0.003$ & $-0.007$ & $-0.017$ \\
  477.5907 & NI & 7.49 & 1.75 & $-2.304$      &C &  8.711 & 8.722 & 8.721   & $-0.011$ & $-0.010$ & $-0.003$ & $-0.007$ & $-0.017$ \\
  477.5907 & KF & 7.49 & 1.42 & $-2.304$      &C &  8.687 & 8.716 & 8.697   & $-0.029$ & $-0.010$ & $-0.006$ & $-0.013$ & $-0.030$ \\
  477.5907 & NF & 7.49 & 1.43 & $-2.304$      &C &  8.691 & 8.720 & 8.701   & $-0.029$ & $-0.010$ & $-0.006$ & $-0.013$ & $-0.030$ \\
  505.2167 & DI & 7.68 & 4.04 & $-1.303$      &B &  8.428 & 8.429 & 8.402   & $-0.000$ & $ 0.026$ & $-0.007$ & $-0.015$ & $-0.036$ \\
  505.2167 & NI & 7.68 & 4.15 & $-1.303$      &B &  8.449 & 8.449 & 8.422   & $ 0.000$ & $ 0.028$ & $-0.007$ & $-0.015$ & $-0.036$ \\
  505.2167 & KF & 7.68 & 3.56 & $-1.303$      &B &  8.456 & 8.482 & 8.442   & $-0.026$ & $ 0.014$ & $-0.013$ & $-0.026$ & $-0.065$ \\
  505.2167 & NF & 7.68 & 3.54 & $-1.303$      &B &  8.452 & 8.478 & 8.438   & $-0.026$ & $ 0.014$ & $-0.013$ & $-0.026$ & $-0.065$ \\
  538.0336 & DI & 7.68 & 2.55 & $-1.616$      &B &  8.444 & 8.447 & 8.428   & $-0.004$ & $ 0.016$ & $-0.007$ & $-0.014$ & $-0.034$ \\
  538.0336 & NI & 7.68 & 2.53 & $-1.616$      &B &  8.439 & 8.442 & 8.423   & $-0.004$ & $ 0.016$ & $-0.007$ & $-0.014$ & $-0.034$ \\
  538.0336 & KF & 7.68 & 2.19 & $-1.616$      &B &  8.465 & 8.490 & 8.458   & $-0.025$ & $ 0.007$ & $-0.012$ & $-0.025$ & $-0.060$ \\
  538.0336 & NF & 7.68 & 2.20 & $-1.616$      &B &  8.468 & 8.492 & 8.461   & $-0.025$ & $ 0.007$ & $-0.012$ & $-0.025$ & $-0.060$ \\
  658.7608 & DI & 8.54 & 1.65 & $-1.003$      &B &  8.316 & 8.322 & 8.293   & $-0.006$ & $ 0.023$ & $-0.004$ & $-0.009$ & $-0.018$ \\
  658.7608 & NI & 8.54 & 1.68 & $-1.003$      &B &  8.326 & 8.332 & 8.303   & $-0.006$ & $ 0.023$ & $-0.004$ & $-0.009$ & $-0.018$ \\
  658.7608 & KF & 8.54 & 1.34 & $-1.003$      &B &  8.343 & 8.369 & 8.336   & $-0.026$ & $ 0.007$ & $-0.007$ & $-0.016$ & $-0.030$ \\
  658.7608 & NF & 8.54 & 1.33 & $-1.003$      &B &  8.339 & 8.365 & 8.332   & $-0.026$ & $ 0.007$ & $-0.007$ & $-0.016$ & $-0.030$ \\
  708.7827 & DI & 8.65 & 0.72 & $-1.442$      &C &  8.413 & 8.423 & 8.401   & $-0.010$ & $ 0.012$ & $-0.004$ & $-0.008$ & $-0.017$ \\
  711.1475 & DI & 8.64 & 1.30 & $-1.085$      &B &  8.348 & 8.355 & 8.328   & $-0.006$ & $ 0.021$ & $-0.004$ & $-0.009$ & $-0.020$ \\
  711.3180 & DI & 8.65 & 2.73 & $-0.773$      &B &  8.468 & 8.467 & 8.430   & $ 0.001$ & $ 0.039$ & $-0.005$ & $-0.011$ & $-0.026$ \\
  711.3180 & NI & 8.65 & 2.75 & $-0.773$      &B &  8.473 & 8.472 & 8.434   & $ 0.001$ & $ 0.039$ & $-0.005$ & $-0.011$ & $-0.026$ \\
  711.3180 & KF & 8.65 & 2.45 & $-0.773$      &B &  8.563 & 8.582 & 8.544   & $-0.019$ & $ 0.018$ & $-0.009$ & $-0.020$ & $-0.045$ \\
  711.3180 & NF & 8.65 & 2.39 & $-0.773$      &B &  8.547 & 8.566 & 8.529   & $-0.019$ & $ 0.018$ & $-0.009$ & $-0.020$ & $-0.045$ \\
  783.7105 & DI & 8.85 & 0.22 & $-1.778$      &B &  8.363 & 8.373 & 8.353   & $-0.010$ & $ 0.010$ & $-0.003$ & $-0.006$ & $-0.012$ \\
  801.8564 & DI & 8.85 & 0.13 & $-2.130$      &D &  8.480 & 8.489 & 8.470   & $-0.009$ & $ 0.010$ & $-0.002$ & $-0.005$ & $-0.011$ \\
  833.5149 & DI & 7.68 &12.30 & $-0.437$      &B+&  8.606 & 8.570 & 8.508   & $ 0.037$ & $ 0.098$ & $-0.069$ & $-0.121$ & $-0.225$ \\
  833.5149 & NI & 7.68 &12.70 & $-0.437$      &B+&  8.642 & 8.605 & 8.542   & $ 0.037$ & $ 0.100$ & $-0.069$ & $-0.121$ & $-0.225$ \\
  833.5149 & KF & 7.68 &10.70 & $-0.437$      &B+&  8.661 & 8.646 & 8.602   & $ 0.014$ & $ 0.058$ & $-0.108$ & $-0.185$ & $-0.318$ \\
  833.5149 & NF & 7.68 &10.60 & $-0.437$      &B+&  8.650 & 8.636 & 8.592   & $ 0.014$ & $ 0.058$ & $-0.108$ & $-0.185$ & $-0.318$ \\
  906.1432 & DI & 7.48 &16.50 & $-0.347$      &B &  8.643 & 8.597 & 8.532   & $ 0.045$ & $ 0.111$ & $-0.074$ & $-0.127$ & $-0.236$ \\
  907.8278 & DI & 7.48 &13.50 & $-0.581$      &B &  8.640 & 8.597 & 8.536   & $ 0.043$ & $ 0.104$ & $-0.059$ & $-0.104$ & $-0.201$ \\
  911.1797 & DI & 7.49 &17.50 & $-0.297$      &B &  8.665 & 8.620 & 8.553   & $ 0.045$ & $ 0.112$ & $-0.054$ & $-0.109$ & $-0.223$ \\
  911.1797 & NI & 7.49 &17.70 & $-0.297$      &B &  8.678 & 8.633 & 8.566   & $ 0.045$ & $ 0.112$ & $-0.054$ & $-0.109$ & $-0.223$ \\
  911.1797 & KF & 7.49 &15.20 & $-0.297$      &B &  8.715 & 8.690 & 8.642   & $ 0.025$ & $ 0.074$ & $-0.146$ & $-0.231$ & $-0.384$ \\
  911.1797 & NF & 7.49 &15.50 & $-0.297$      &B &  8.739 & 8.714 & 8.664   & $ 0.025$ & $ 0.075$ & $-0.146$ & $-0.231$ & $-0.384$ \\
  960.3032 & DI & 7.48 &11.50 & $-0.896$      &B &  8.727 & 8.686 & 8.630   & $ 0.041$ & $ 0.098$ & $-0.039$ & $-0.073$ & $-0.130$ \\
  965.8435 & DI & 7.49 &17.00 & $-0.280$      &B &  8.559 & 8.512 & 8.447   & $ 0.047$ & $ 0.112$ & $-0.042$ & $-0.094$ & $-0.179$ \\
 1012.3871 & DI & 8.54 &11.80 & $-0.031$      &C+&  8.682 & 8.651 & 8.589   & $ 0.031$ & $ 0.094$ & $-0.032$ & $-0.063$ & $-0.118$ \\
 1012.3871 & NI & 8.54 &11.60 & $-0.031$      &C+&  8.665 & 8.634 & 8.572   & $ 0.031$ & $ 0.093$ & $-0.032$ & $-0.063$ & $-0.118$ \\
 1012.3871 & KF & 8.54 & 9.41 & $-0.031$      &C+&  8.677 & 8.672 & 8.626   & $ 0.006$ & $ 0.051$ & $-0.056$ & $-0.106$ & $-0.170$ \\
 1012.3871 & NF & 8.54 & 9.35 & $-0.031$      &C+&  8.671 & 8.665 & 8.620   & $ 0.006$ & $ 0.051$ & $-0.056$ & $-0.106$ & $-0.170$ \\
 1054.1241 & DI & 8.54 & 2.03 & $-1.398$      &D &  8.728 & 8.725 & 8.699   & $ 0.004$ & $ 0.030$ & $-0.009$ & $-0.017$ & $-0.042$ \\
 1054.1241 & NI & 8.54 & 1.95 & $-1.398$      &D &  8.706 & 8.703 & 8.677   & $ 0.003$ & $ 0.029$ & $-0.009$ & $-0.017$ & $-0.042$ \\
 1054.1241 & KF & 8.54 & 1.75 & $-1.398$      &D &  8.800 & 8.814 & 8.788   & $-0.014$ & $ 0.011$ & $-0.016$ & $-0.032$ & $-0.074$ \\
 1054.1241 & NF & 8.54 & 1.65 & $-1.398$      &D &  8.767 & 8.781 & 8.756   & $-0.015$ & $ 0.010$ & $-0.016$ & $-0.032$ & $-0.074$ \\
 1068.5345 & DI & 7.48 &20.20 & $-0.272$      &B &  8.605 & 8.556 & 8.489   & $ 0.050$ & $ 0.116$ & $-0.067$ & $-0.117$ & $-0.205$ \\
 1068.5345 & NI & 7.48 &20.00 & $-0.272$      &B &  8.594 & 8.544 & 8.478   & $ 0.050$ & $ 0.116$ & $-0.067$ & $-0.117$ & $-0.205$ \\
 1068.5345 & KF & 7.48 &17.80 & $-0.272$      &B &  8.668 & 8.640 & 8.588   & $ 0.028$ & $ 0.080$ & $-0.107$ & $-0.188$ & $-0.340$ \\
 1068.5345 & NF & 7.48 &17.50 & $-0.272$      &B &  8.648 & 8.620 & 8.569   & $ 0.028$ & $ 0.080$ & $-0.107$ & $-0.188$ & $-0.340$ \\
 1070.7333 & DI & 7.48 &18.50 & $-0.411$      &B &  8.638 & 8.589 & 8.524   & $ 0.050$ & $ 0.114$ & $-0.047$ & $-0.090$ & $-0.193$ \\
 1070.7333 & NI & 7.48 &18.20 & $-0.411$      &B &  8.619 & 8.570 & 8.505   & $ 0.050$ & $ 0.114$ & $-0.047$ & $-0.090$ & $-0.193$ \\
 1070.7333 & KF & 7.48 &16.80 & $-0.411$      &B &  8.735 & 8.707 & 8.656   & $ 0.028$ & $ 0.078$ & $-0.127$ & $-0.207$ & $-0.357$ \\
 1070.7333 & NF & 7.48 &16.00 & $-0.411$      &B &  8.678 & 8.651 & 8.601   & $ 0.027$ & $ 0.077$ & $-0.127$ & $-0.207$ & $-0.357$ \\
 1072.9533 & DI & 7.49 &18.20 & $-0.420$      &B &  8.644 & 8.594 & 8.529   & $ 0.050$ & $ 0.115$ & $-0.040$ & $-0.089$ & $-0.190$ \\
 1072.9533 & NI & 7.49 &17.80 & $-0.420$      &B &  8.618 & 8.568 & 8.504   & $ 0.050$ & $ 0.115$ & $-0.040$ & $-0.089$ & $-0.190$ \\
 1072.9533 & KF & 7.49 &15.80 & $-0.420$      &B &  8.687 & 8.660 & 8.611   & $ 0.028$ & $ 0.077$ & $-0.120$ & $-0.199$ & $-0.342$ \\
 1072.9533 & NF & 7.49 &15.60 & $-0.420$      &B &  8.672 & 8.645 & 8.596   & $ 0.027$ & $ 0.076$ & $-0.120$ & $-0.199$ & $-0.342$ \\
 1075.3985 & DI & 7.49 & 4.54 & $-1.606$      &B &  8.489 & 8.467 & 8.435   & $ 0.022$ & $ 0.055$ & $-0.012$ & $-0.025$ & $-0.055$ \\
 1075.3985 & NI & 7.49 & 4.48 & $-1.606$      &B &  8.480 & 8.458 & 8.426   & $ 0.022$ & $ 0.054$ & $-0.012$ & $-0.025$ & $-0.055$ \\
 1075.3985 & KF & 7.49 & 3.64 & $-1.606$      &B &  8.477 & 8.478 & 8.454   & $-0.001$ & $ 0.023$ & $-0.023$ & $-0.045$ & $-0.095$ \\
 1075.3985 & NF & 7.49 & 3.61 & $-1.606$      &B &  8.471 & 8.472 & 8.449   & $-0.001$ & $ 0.023$ & $-0.023$ & $-0.045$ & $-0.095$ \\
 1174.8220 & DI & 8.64 &14.90 & $ 0.375$      &B &  8.368 & 8.330 & 8.267   & $ 0.038$ & $ 0.102$ & $-0.029$ & $-0.060$ & $-0.138$ \\
 1174.8220 & NI & 8.64 &14.80 & $ 0.375$      &B &  8.362 & 8.323 & 8.260   & $ 0.038$ & $ 0.101$ & $-0.029$ & $-0.060$ & $-0.138$ \\
 1174.8220 & NF & 8.64 &12.40 & $ 0.375$      &B &  8.392 & 8.381 & 8.327   & $ 0.011$ & $ 0.064$ & $-0.032$ & $-0.086$ & $-0.211$ \\
 1177.7540 & DI & 8.64 & 6.56 & $-0.520$      &B &  8.514 & 8.488 & 8.443   & $ 0.026$ & $ 0.071$ & $-0.015$ & $-0.030$ & $-0.073$ \\
 1177.7540 & NI & 8.64 & 6.58 & $-0.520$      &B &  8.517 & 8.491 & 8.446   & $ 0.026$ & $ 0.071$ & $-0.015$ & $-0.030$ & $-0.073$ \\
 1177.7540 & KF & 8.64 & 5.09 & $-0.520$      &B &  8.491 & 8.492 & 8.455   & $-0.002$ & $ 0.036$ & $-0.026$ & $-0.053$ & $-0.125$ \\
 1177.7540 & NF & 8.64 & 5.09 & $-0.520$      &B &  8.491 & 8.492 & 8.455   & $-0.002$ & $ 0.036$ & $-0.026$ & $-0.053$ & $-0.125$ \\
 1184.8730 & DI & 8.64 & 5.70 & $-0.697$      &B &  8.510 & 8.490 & 8.452   & $ 0.020$ & $ 0.058$ & $-0.012$ & $-0.025$ & $-0.060$ \\
 1184.8730 & NI & 8.64 & 5.75 & $-0.697$      &B &  8.516 & 8.496 & 8.458   & $ 0.020$ & $ 0.058$ & $-0.012$ & $-0.025$ & $-0.060$ \\
 1184.8730 & KF & 8.64 & 4.66 & $-0.697$      &B &  8.534 & 8.538 & 8.502   & $-0.004$ & $ 0.032$ & $-0.023$ & $-0.049$ & $-0.108$ \\
 1186.2990 & DI & 8.64 & 5.81 & $-0.710$      &B &  8.535 & 8.514 & 8.476   & $ 0.020$ & $ 0.059$ & $-0.011$ & $-0.025$ & $-0.061$ \\
 1186.2990 & NI & 8.64 & 5.79 & $-0.710$      &B &  8.532 & 8.512 & 8.474   & $ 0.020$ & $ 0.058$ & $-0.011$ & $-0.025$ & $-0.061$ \\
 1186.2990 & KF & 8.64 & 4.63 & $-0.710$      &B &  8.541 & 8.544 & 8.508   & $-0.004$ & $ 0.032$ & $-0.023$ & $-0.049$ & $-0.107$ \\
 1189.2910 & DI & 8.64 &10.10 & $-0.277$      &B &  8.518 & 8.491 & 8.439   & $ 0.027$ & $ 0.079$ & $-0.020$ & $-0.042$ & $-0.097$ \\
 1189.2910 & NI & 8.64 &10.10 & $-0.277$      &B &  8.518 & 8.491 & 8.439   & $ 0.027$ & $ 0.079$ & $-0.020$ & $-0.042$ & $-0.097$ \\
 1189.2910 & NF & 8.64 & 8.45 & $-0.277$      &B &  8.555 & 8.553 & 8.506   & $ 0.002$ & $ 0.048$ & $-0.037$ & $-0.076$ & $-0.157$ \\
 1189.5750 & DI & 8.65 &13.94 & $-0.008$      &B &  8.530 & 8.499 & 8.441   & $ 0.031$ & $ 0.089$ & $-0.027$ & $-0.052$ & $-0.117$ \\
 1189.5750 & NI & 8.65 &13.97 & $-0.008$      &B &  8.532 & 8.501 & 8.442   & $ 0.031$ & $ 0.090$ & $-0.027$ & $-0.052$ & $-0.117$ \\
 1189.5750 & NF & 8.65 &11.45 & $-0.008$      &B &  8.550 & 8.545 & 8.492   & $ 0.005$ & $ 0.058$ & $-0.023$ & $-0.073$ & $-0.169$ \\
 1254.9480 & DI & 8.85 & 6.80 & $-0.565$      &B &  8.632 & 8.604 & 8.563   & $ 0.028$ & $ 0.069$ & $-0.008$ & $-0.017$ & $-0.045$ \\
 1256.2120 & DI & 8.85 & 6.89 & $-0.522$      &B &  8.597 & 8.569 & 8.528   & $ 0.028$ & $ 0.069$ & $-0.008$ & $-0.018$ & $-0.047$ \\
 1256.9040 & DI & 8.85 & 5.75 & $-0.598$      &B &  8.535 & 8.510 & 8.473   & $ 0.025$ & $ 0.062$ & $-0.008$ & $-0.017$ & $-0.044$ \\
 1258.1590 & DI & 8.85 & 6.00 & $-0.536$      &B &  8.502 & 8.477 & 8.439   & $ 0.026$ & $ 0.063$ & $-0.008$ & $-0.018$ & $-0.042$ \\
 1734.6381 & DI & 9.70 & 0.64 & $-1.348$      &C &  8.352 & 8.336 & 8.351   & $ 0.016$ & $ 0.001$ & $ 0.000$ & $ 0.000$ & $ 0.003$ \\
 1744.8600 & DI & 9.00 &21.60 & $ 0.012$      &B+&  8.428 & 8.393 & 8.359   & $ 0.035$ & $ 0.069$ & $-0.006$ & $-0.015$ & $-0.058$ \\
 1745.5971 & DI & 9.70 &17.40 & $ 0.280$      &C &  8.567 & 8.534 & 8.502   & $ 0.033$ & $ 0.066$ & $ 0.001$ & $ 0.002$ & $ 0.008$ \\
 1750.5641 & DI & 9.70 &17.50 & $ 0.424$      &C &  8.430 & 8.396 & 8.364   & $ 0.034$ & $ 0.066$ & $ 0.001$ & $ 0.001$ & $ 0.008$ \\
 1763.7381 & DI & 9.71 &21.00 & $ 0.338$      &C &  8.673 & 8.639 & 8.600   & $ 0.035$ & $ 0.073$ & $ 0.001$ & $ 0.002$ & $ 0.013$ \\
 1778.9600 & DI & 7.95 & 1.40 & $-2.246$      &B &  8.247 & 8.225 & 8.239   & $ 0.022$ & $ 0.009$ & $-0.003$ & $-0.006$ & $-0.008$ \\
 2102.3131 & DI & 9.17 &10.00 & $-0.450^{~b}$ &--&  8.651 & 8.618 & 8.580   & $ 0.033$ & $ 0.071$ & $-0.001$ & $-0.002$ & $-0.009$ \\
 2121.1551 & DI & 9.83 & 8.56 & $-0.080^{~b}$ &--&  8.660 & 8.632 & 8.599   & $ 0.027$ & $ 0.061$ & $ 0.000$ & $ 0.000$ & $ 0.004$ \\
 2125.9891 & DI & 9.83 &17.80 & $ 0.490^{~b}$ &--&  8.598 & 8.566 & 8.513   & $ 0.033$ & $ 0.085$ & $ 0.000$ & $ 0.000$ & $ 0.004$ \\
 3129.7480 & DI & 9.69 & 2.60 & $-0.570^{~b}$ &--&  8.688 & 8.673 & 8.650   & $ 0.015$ & $ 0.039$ & $ 0.000$ & $ 0.001$ & $ 0.002$ \\
 2290.6561 & DI & 9.17 &12.15 & $-0.182^{~a}$ &--&  8.793 & 8.763 & 8.720   & $ 0.030$ & $ 0.073$ & $-0.001$ & $-0.003$ & $-0.013$ \\
\noalign{\smallskip}
\hline
\end{longtable}
\loggf-values with their quality (column six) are taken from the NIST database;
the quality (Acc.) represents the accuracy of the value, ranging from B+, meaning 
${\sigma f\over f}\le 7\%$, to  D, meaning ${\sigma f\over f}\le 50\%$. \loggf\ with 
flag b are from \citet{biemont93}, those with flag a are from 
\citet{asplund05} (no quality is indicated in these cases).
}
\twocolumn

In Table\,\ref{let_ac} the carbon abundance determinations in the last thirty
years are listed. The difference of 0.28\,dex from the highest to the lowest
value cannot be explained with NLTE effects, that, according to our analysis,
are about --0.05\,dex on average for ${\rm S_H}=1/3$.
The average of the values in the Table is 8.51 with a standard deviation
of 0.09.

\begin{table}
\caption{Solar abundance of C in the literature}
\label{let_ac}
\begin{center}
\begin{tabular}{ll}
\noalign{\smallskip}\hline\noalign{\smallskip}
A(C)      & Ref   \\ 
\noalign{\smallskip}\hline\noalign{\smallskip}
8.67~ $\pm$0.10  & \citet{lambert78}  \\ 
8.56~ $\pm$0.04  & \citet{ag89}       \\ 
8.58~ $\pm$0.13  & \citet{stuerenburg}\\ 
8.60~ $\pm$0.05  & \citet{grevesse91} \\ 
8.60~ $\pm$0.10  & \citet{biemont93}  \\ 
8.55             & \citet{grevesse94} \\ 
8.54             & \citet{takeda94}   \\ 
8.52~ $\pm$0.06  & \citet{grevesse98} \\ 
8.52~ $\pm$0.06  & \citet{grevesse00} \\ 
8.57~ $\pm$0.03  & \citet{hhoxy}      \\ 
8.592$\pm$0108   & \citet{hhoxy}      \\ 
8.39~ $\pm$0.04  & \citet{allende02}  \\ 
8.39~ $\pm$0.05  & \citet{asplund05}  \\ 
8.39~ $\pm$0.05  & \citet{scott06}    \\ 
8.44~ $\pm$0.06  & \citet{pin06}      \\ 
8.39~ $\pm$0.05  & \citet{grevesse07} \\ 
8.43~ $\pm$0.05  & \citet{asplund09}  \\ 
8.50~ $\pm$0.06  & present work       \\
\noalign{\smallskip}\hline\noalign{\smallskip}
\end{tabular}
\\
\end{center}
\end{table}

\section{Conclusions}

Our recommended value for the solar carbon abundance
is {\bf A(C) = $ 8.50\pm 0.06$,} corresponding
to a weak efficiency of the collisions with 
neutral hydrogen atoms (${\rm S_{\rm H}}=1/3$),
the favourite value of the Holweger school.
The quoted error is the linear sum
of a statistical error, 0.02\, dex, and
a systematic error, 0.04\,dex, due to the
uncertainty of the treatment of the hydrogen 
collisions in the NLTE computation.
The statistical error has been estimated by dividing
the line-to-line scatter, 0.11\, dex, by
the square root of the 45 independent lines
used in the analysis. 
The value ${\rm S_{\rm H}}=1/3$ was also
adopted in our investigations of the solar abundances
of oxygen \citep{oxy} and nitrogen
\citep{nitro}. It is not obvious 
why the same value of ${\rm S_{\rm H}}$ should
apply to different atoms, or even to 
different lines of the same atom.
It is comforting that the difference between
the extreme assumptions about the efficiency of
the H collisions (${\rm S_{\rm H}}=0$ or $1$)
amounts to only 0.08\, dex.

Our preferred value for the
solar carbon abundance is very close to the
recommendation of \citet{grevesse98}.
If we take this carbon abundance, together with
A(N)=7.86 from \citet{nitro}, 
A(O)=8.76 from \citet{oxy} and
A(Ne)=8.02 (see \citealt{nitro}
for an explanation of this choice),
we obtain a solar metallicity of $Z=0.0154$
and $Z/X = 0.0211$. 
This value is higher than the metallicity recommended
by \citet{sunabboasp} and \citet{grevesse07}, $Z=0.0122$,
and goes in the direction of reconciling the spectroscopic 
abundances with the constraints from helioseismology.

The fact that different 3D hydrodynamical
simulations provide significantly different
results (of the order of 0.1\, dex)
underlines the need for further development and 
validation of the hydrodynamical models.
Recently \citet{asplund09} has presented results based
on a new generation 3D model, which is much closer
to the \cobold\ model than the one used by 
\citet{asplund05}, and lead to an upward
revision of his carbon, oxygen, and iron abundances.
It is likely that any residual difference between this
new result and the present analysis can be ascribed
to the line selection and to the different treatment 
of hydrogen collisions in the NLTE computations, although 
details on the analysis of \citet{asplund09} are not yet 
available. The excellent agreement of our \cobold\ solar 
model with the observed centre-to-limb variation, shown in
\citet{ludwig09}, provides a strong support to the
thermal structure of the model and to the abundances
deduced by its application \citep{caffau09a}.

\begin{acknowledgements}
EC,HGL and PB acknowledge financial support
from EU contract MEXT-CT-2004-014265 (CIFIST).
\end{acknowledgements}



\begin{thebibliography}{}

\bibitem[Allende Prieto et al.(2002)]{allende02} Allende Prieto, 
C., Lambert, D.~L., \& Asplund, M.\ 2002, \apjl, 573, L137

\bibitem[Anders \& Grevesse(1989)]{ag89} Anders, E., \& Grevesse, N.\ 1989, \gca, 53, 197

\bibitem[Asplund et 
al.(2004)]{asplund04} Asplund, M., Grevesse, N., Sauval, A.~J., Allende Prieto, C., \& Kiselman, D.\ 2004, \aap, 417, 751 

\bibitem[Asplund et 
al.(2005a)]{asplund05} Asplund, M., Grevesse, N., Sauval, A.~J., Allende Prieto, C., \& Blomme, R.\ 2005a, \aap, 431, 693

\bibitem[Asplund et al. (2005b)]{sunabboasp} Asplund, M., Grevesse, 
N., \& Sauval, A.~J.\ 2005b, ASP Conf.~Ser.~336: Cosmic Abundances as 
Records of Stellar Evolution and Nucleosynthesis, 336, 25 

\bibitem[Asplund et 
al.(2009)]{asplund09} Asplund, M., Grevesse, N., Sauval, A.~J., \& Scott, P.\ 2009, \araa, 47, 481

\bibitem[Ayres et al.(2006)]{ayres06} Ayres, T.~R., Plymate, 
C., \& Keller, C.~U.\ 2006, \apjs, 165, 618

\bibitem[Barklem et al.(1998)]{abo4} Barklem, P.~S., Anstee, 
S.~D., \& O'Mara, B.~J.\ 1998, Publications of the Astronomical Society of 
Australia, 15, 336

\bibitem[Baschek 
\& Holweger(1967)]{baschek67} Baschek, B., \& Holweger, H.\ 1967, Zeitschrift fur Astrophysik, 67, 143 

\bibitem[Basu \& Antia(2008)]{basu} Basu, S., \& Antia, H.~M.\ 2008, \physrep, 457, 217

\bibitem[Bi{\`e}mont et al.(1993)]{biemont93} Bi{\`e}mont, E., Hibbert, 
A., Godefroid, M., \& Vaeck, N.\ 1993, \apj, 412, 431

\bibitem[Caffau \& Ludwig(2007)]{zolfito} Caffau, E., \&
Ludwig, H.-G.\ 2007, \aap, 467, L11

\bibitem[Caffau et al.(2009a)]{caffau09a}
Caffau, E., Ludwig,H.-G., Steffen, M.\ 2009a, MmSAI, Vol.~80, n.~3

\bibitem[Caffau et al.(2008)]{oxy} Caffau, E., Ludwig, 
H.-G., Steffen, M., Ayres, T.~R., Bonifacio, P., Cayrel, R., Freytag, B., 
\& Plez, B.\ 2008, A\&A, 488, 1031 

\bibitem[Caffau et al.(2009b)]{nitro} Caffau, E., Maiorca, E., 
Bonifacio, P., Faraggiana R., Steffen, M., H.-G. Ludwig, 
Kamp, I., Busso, M.\ 2009b, \aap, 498, 877

\bibitem[Castelli \& Kurucz(2003)]{ck03} Castelli, F., \& Kurucz, R.~L.\ 2003,
in Modelling of Stellar Atmospheres, IAU Symp. No. 210,
eds. N. Piskunov et al., Poster A20, arXiv:astro-ph/0405087

\bibitem[Chaplin 
\& Basu(2008)]{chaplin08} Chaplin, W.~J., \& Basu, S.\ 2008, \solphys, 251, 53 

\bibitem[Delahaye 
\& Pinsonneault(2006)]{delahaye06} Delahaye, F., \& Pinsonneault, M.~H.\ 2006, \apj, 649, 529

\bibitem[Delbouille et al.(1973)]{delbouille} Delbouille, L., 
Roland, G., \& Neven, L.\ 1973, Liege: Universite de Liege, Institut 
d'Astrophysique, 1973

\bibitem[Delbouille et al.(1981)]{delbouilleir}Delbouille L., Roland G., Brault, Testerman 1981;
``Photometric atlas of the solar spectrum from 1850 to 10,000 cm$^{-1}$'',
http://bass2000.obspm.fr/solar\_spect.php

\bibitem[Drawin(1969)]{Drawin}Drawin, H.W., 1969, Z. Physik 225, 483

\bibitem[Farmer(1994)]{farmer} Farmer, C.~B.\ 1994, Infrared 
Solar Physics, 154, 511 

\bibitem[Farmer et al.(1989)]{atmos} Farmer, C.~B., Norton, 
R.~H., \& Geller, M.\ 1989, NASA Reference Publication, 1224,  

\bibitem[{{Freytag} {et~al.}(2002), {Steffen}, \&
  {Dorch}}]{freytag02}
{Freytag}, B., {Steffen}, M., \& {Dorch}, B. 2002, Astronomische Nachrichten,
  323, 213

\bibitem[{{Freytag} {et~al.}(2003){Freytag}, {Steffen}, {Wedemeyer-B{\"o}hm},
  \& {Ludwig}}]{freytag03}
{Freytag}, B., {Steffen}, M., {Wedemeyer-B{\"o}hm}, S., \& {Ludwig}, H.-G.
  2003, {CO5BOLD User Manual},
  \verb|http://www.astro.uu.se/~bf/co5bold_main.html|

\bibitem[Grevesse et al.(1987)]{grevesse87} Grevesse, N., Sauval, 
A.~J., Farmer, C.~B., 
\& Norton, R.~H.\ 1987, Liege International Astrophysical Colloquia, 27, 111 

\bibitem[Grevesse et 
al.(1991)]{grevesse91} Grevesse, N., Lambert, D.~L., Sauval, A.~J., van Dishoek, E.~F., Farmer, C.~B., \& Norton, R.~H.\ 1991, \aap, 242, 488

\bibitem[Grevesse et al.(1994)]{grevesse94} Grevesse, N., Sauval, 
A.~J., \& Blomme, R.\ 1994, Infrared Solar Physics, 154, 539 

\bibitem[Grevesse 
\& Sauval(1998)]{grevesse98} Grevesse, N., \& Sauval, A.~J.\ 1998, Space Science Reviews, 85, 161

\bibitem[Grevesse et al.(2000)]{grevesse00} Grevesse, N., Sauval, 
A., \& Murdin, P.\ 2000, Encyclopedia of Astronomy and Astrophysics

\bibitem[Grevesse et al.(2007)]{grevesse07} Grevesse, N., Asplund, 
M., \& Sauval, A.~J.\ 2007, Space Science Reviews, 130, 105

\bibitem[Hibbert et 
al.(1993)]{hibbert93} Hibbert, A., Biemont, E., Godefroid, M., \& Vaeck, N.\ 1993, \aaps, 99, 179

\bibitem[Holweger(1967)]{hhsunmod} Holweger, H.\ 1967,
Zeitschrift fur Astrophysik, 65, 365

\bibitem[Holweger(2001)]{hhoxy} Holweger, H.\ 2001, AIP
Conf.~Proc.~598: Joint SOHO/ACE workshop ''Solar and Galactic 
Composition'', 598, 23

\bibitem[Holweger \& M\"uller(1974)]{hmsunmod} Holweger, H., \&
M\"uller, E.~A.\ 1974, \solphys, 39, 19

\bibitem[Kurucz(2005a)]{kuruczflux} Kurucz, R.~L.\ 2005a, Memorie 
della Societ\`a Astronomica Italiana Supplementi, 8, 189

\bibitem[Lambert(1978)]{lambert78} Lambert, D.~L.\ 1978, \mnras, 
182, 249 

\bibitem[Ludwig et al. (2009)]{ludwig09}
Ludwig, H.-G., Caffau, E.,  Bonifacio, P.,
Steffen, M.,  Freytag, B., Cayrel, R.\ 2009
in IAU Symposium 265, K. Cunha, M. Spite \& B. Barbuy eds., 
p. 

\bibitem[Luo 
\& Pradhan(1989)]{luo89} Luo, D., \& Pradhan, A.~K.\ 1989, Journal of Physics B Atomic Molecular Physics, 22, 3377

\bibitem[Neckel \& Labs(1984)]{neckelobs} Neckel, H., \& Labs, 
D.\ 1984, \solphys, 90, 205 

\bibitem[Neckel(1999)]{neckel1999} Neckel, H.\ 1999, \solphys, 184, 421

\bibitem[Nussbaumer 
\& Storey(1984)]{nussbaumer84} Nussbaumer, H., \& Storey, P.~J.\ 1984, \aap, 140, 383

\bibitem[Pinsonneault \& Delahaye(2006)]{pin06} Pinsonneault, M.~H., \& 
Delahaye, F.\ 2006, ApJ submitted, arXiv:astro-ph/0606077

\bibitem[Ralchenko(2005)]{ralchenko} Ralchenko, Y.\ 2005, Memorie 
della Societ\`a Astronomica Italiana Supplementi, 8, 96

\bibitem[Ryan(1998)]{ryan} Ryan, S.~G.\ 1998, \aap, 331, 1051

\bibitem[Scott et 
al.(2006)]{scott06} Scott, P.~C., Asplund, M., Grevesse, N., \& Sauval, A.~J.\ 2006, \aap, 456, 675 


\bibitem[Steenbock \& Holweger(1984)]{SH} Steenbock, W.,
\& Holweger, H.\ 1984, \aap, 130, 319

\bibitem[Steffen 
\& Holweger(2002)]{mst02} Steffen, M., \& Holweger, H.\ 2002, \aap, 387, 258

\bibitem[St{\"u}renburg 
\& Holweger(1990)]{stuerenburg} St{\"u}renburg, S., \& Holweger, H.\ 1990, \aap, 237, 125

\bibitem[Takeda(1994)]{takeda94} Takeda, Y.\ 1994, \pasj, 46, 53

\bibitem[Tody(1993)]{tody} Tody, D.\ 1993, Astronomical Data 
Analysis Software and Systems II, 52, 173 

\bibitem[{{Wedemeyer} {et~al.}(2004){Wedemeyer}, {Freytag}, {Steffen},
  {Ludwig}, \& {Holweger}}]{Wedemeyer}
{Wedemeyer}, S., {Freytag}, B., {Steffen}, M., {Ludwig}, H.-G., \& {Holweger},
  H. 2004, \aap, 414, 1121

\bibitem[Weiss(1996)]{weiss96}Weiss, A. W.\ 1996, private communication to NIST

\bibitem[Wiese et al.(1996)]{wiese96} Wiese, W.~L., Fuhr, 
J.~R., 
\& Deters, T.~M.\ 1996, Atomic transition probabilities of carbon, nitrogen, and oxygen: 
a critical data compilation.~ Edited by W.L.~Wiese, J.R.~Fuhr, and T.M.~Deters.~Washington, DC:  
American Chemical Society ...~for the National Institute of Standards and Technology (NIST) c1996.~QC 453 .W53 1996

\bibitem[Yang 
\& Bi(2007)]{yang07} Yang, W.~M., \& Bi, S.~L.\ 2007, \apjl, 658, L67
\end{thebibliography}
\end{document}